\documentclass[10pt]{article}

\usepackage{amsmath}
\usepackage{amssymb}

\usepackage{graphicx}

\usepackage{cite}
\usepackage{color}
\usepackage{booktabs,colortbl}

\usepackage[labelfont=bf,labelsep=period,justification=raggedright]{caption}
\usepackage[colorlinks,linkcolor=blue,anchorcolor=blue,citecolor=blue]{hyperref}

\makeatletter
\renewcommand{\@biblabel}[1]{\quad#1.}
\makeatother

\date{}

\begin{document}

\begin{flushleft}
{\Large
\textbf{Universal underpinning of human mobility in the real world and
cyberspace}
}
\\
Yi-Ming Zhao$^1$,
An Zeng$^{1\star}$,
Xiao-Yong Yan$^{2\dagger}$,
Wen-Xu Wang$^1$,
Ying-Cheng Lai$^3$

1, School of Systems Science, Beijing Normal University,
Beijing, 100875, P. R. China\\
2, Systems Science Institute, Beijing Jiaotong University, Beijing 100044, P. R. China\\
3, School of Electrical, Computer and Energy Engineering, Arizona State University, Tempe, Arizona 85287, USA\\
$\star$ corresponding author: anzeng@bnu.edu.cn\\
$\dagger$ corresponding author: kaiseryxy@163.com\\








\end{flushleft}

\section*{Abstract}
Human movements in the real world and in cyberspace
affect not only dynamical processes such as epidemic spreading and
information diffusion but also social and economical activities such as
urban planning and personalized recommendation in online shopping. Despite
recent efforts in characterizing and modeling human behaviors in both the
real and cyber worlds, the fundamental dynamics underlying human mobility
have not been well understood. We develop a minimal, memory-based random
walk model in limited space for reproducing, with a single parameter,
the key statistical behaviors characterizing human movements in both spaces.
The model is validated using big data from mobile phone and online
commerce, suggesting memory-based random walk dynamics as the universal
underpinning for human mobility, regardless of whether it occurs in the
real world or in cyberspace.

\newpage
\tableofcontents

\section{Introduction}
Human mobility in the real world and cyberspace plays an ever increasing
role in the modern society and economy. Many important processes are
affected by the patterns of human mobility, such as epidemic and information
spreading~\cite{Lima,AA,VB,Plos,Quan},
traffic congestion~\cite{Dyna,Areview,All}, and
e-commerce. Modern research on human
mobility dynamics began with the trajectory-based approach~\cite{Bro},
e.g., by tracing the trajectories of dollar bills in the real world,
which revealed a number of scaling relations such as a truncated
power law in the distribution of the traveling distance. Analysis of
the mobile phone data demonstrated that the individual travel patterns
can be characterized by a spatial probability distribution,
indicating the existence of universal patterns in the human
trajectories~\cite{Mat}. The question of whether human mobility
patterns are predictable was addressed through an analysis of the limits
of predictability in human dynamics~\cite{SCM1}. More recently, human
mobility in the cyberspace and its relation to that in the physical
space were studied using big data analysis and phenomenological
modeling~\cite{ZDZZ}.

Fundamental to the study of human mobility dynamics is the development
of models to reproduce the phenomena and scaling relations obtained from
empirical data~\cite{Asurvey}. A pioneering work in this field is the
articulation of a statistical, self-consistent microscopic
model~\cite{SCM2}. Subsequent studies focused on predicting the
mobility flow between two locations through, e.g., the classic
gravity model~\cite{Zipf}. A stochastic process capturing local
mobility decisions, the so-called radiation model, was
introduced~\cite{Simini}, which yields better agreement with the
empirical data than the gravity model. Alternative mechanisms were
introduced to model the human trajectories~\cite{Mobili,Net,Yana,Yan,LLH,Yanqing,Sara}.
The modeling effort has also been extended to the cyberspace~\cite{ZDZZ}.
While many models were developed to reproduce the scaling laws obtained
from various human mobility empirical data, a physical and first-principle
based understanding of the underlying dynamics is still missing.
In particular, the widely studied model of human microscopic
trajectories~\cite{SCM2} imposes the hypothesis
that the probability for individuals to visit new locations in the physical
space has a power-law form: $P_{new} = \rho S^{-\gamma}$, where $S$ is the
number of distinct locations already visited, with $\rho$ and $\gamma$
being two parameters.
However, the underlying mechanism accounting for the power-law probability of exploring new locations is yet elusive, prompting us to wonder if there is a universal, minimal model capable of predicting all known scaling laws for human mobility in both the real world and cyberspace.

In this paper, we show that all observed statistical features
of human trajectories in the physical space and cyberspace can be
quantitatively predicted through a universal mechanism: memory based,
preferential random walk process. The memory effect has been known to be
important to human dynamics in general~\cite{YY,Catt,Cir,GGG,SSS}, which in limited space is
a basic ingredient in our minimal model. The probability for an individual
to visit a new location can then be obtained from {\em first principle
considerations} without the need to hypothesize any particular mathematical
form. The basic idea is intuitive (as most of us have experience with): if
an individual visited a location in the past, the location would imprint a
memory effect on the individual, enhancing the probability for him/her
to visit the same location in the future. The striking finding is that, this
simple rule, with only a single parameter, is capable of generating all the
known statistical properties of human mobility (e.g., those predicted by
the models of self-consistency~\cite{SCM2} with more parameters and
scaling assumption about the probability). Solving our minimal model
analytically, we obtain scaling relations that agree well with the
empirical ones from mobile phone check-ins and online shopping data sets that
record human trajectories in the real world and cyberspace, respectively.
Our minimal model thus establishes the universal underpinning of human
mobility, representing a significant step forward in understanding modern
human behaviors through statistical physics. This has the potential to
advance a number of disciplines such as social sciences and online economics.

\section{Methods}
\subsection{Memory-preferential random walk model.} We consider a finite space of $M$ locations, in which $N$ individuals
perform random walk with the probability of visiting a position proportional
to its weight. For convenience, we use Latin and Greek letters to denote
individuals and locations, respectively. The weight of a location $\alpha$
with respect to individual $i$, $w_\alpha^i$, is updated during the process.
An actual visit of $i$ to $\alpha$ will increase the weight $w_\alpha^i$
through $\lambda$ - the memory factor parameter. For $\lambda=0$ and
$\lambda>0$, the random walk is unbiased and memory-preferential,
respectively.

\begin{figure}[h]
\centering
\includegraphics[width=0.68\linewidth]{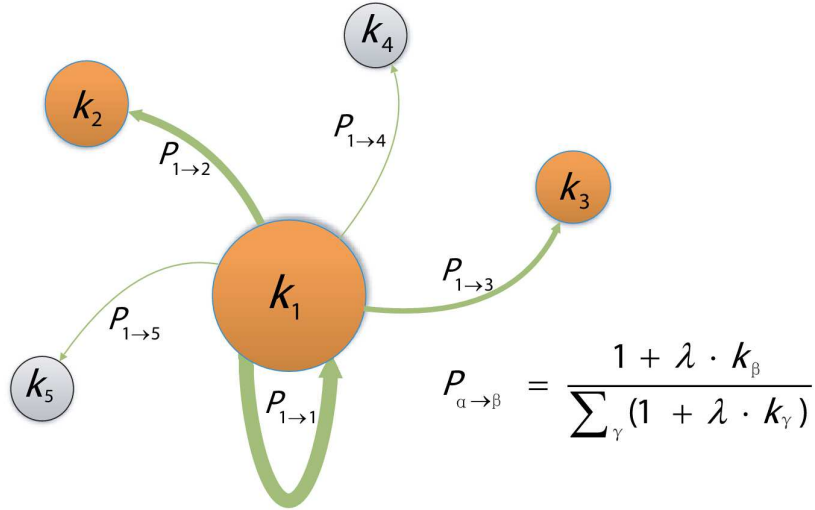}
\caption{\textbf{Illustration of memory-preferential random walk
process.} Initially, every location has unit weight. At each time step,
the walker chooses a location as the next destination. The visited times
to location $\beta$ is denoted as $k_{\beta}$. The probability for
the walker to choose location $\beta$ is propositional to the initial
weight of this location plus the product of the memory factor parameter
$\lambda$ and $k_{\beta}$.}
\label{fig:MPRW}
\end{figure}

In our memory-preferential random walk (MPRW) model, for individual $i$
the weight sequence of all $M$ locations at time step $t$ can be written as
$\{1+\lambda k_{1}^i(t),1+\lambda k_{2}^i(t),1+\lambda k_{3}^i(t),
\ldots,1+\lambda k_{M}^i(t)\}$, where $k_\alpha^i(t)$ is the number of
times that position $\alpha$ has been visited before time $t$. When $i$
is about to move to a new location at $t+1$, the probability to go to
$\alpha$ is proportional to the weight of $\alpha$, i.e.,
$w_\alpha^i \sim 1+\lambda k_\alpha^i(t)$. We have
$$p_{\alpha}^i(t+1) =\frac{1+\lambda k_{\alpha}^i(t)}{\Sigma_{\beta}
[1+\lambda k_{\beta}^i(t)]}.$$
A typical step of the memory-preferential random walk process is
schematically illustrated in Fig.~\ref{fig:MPRW}.

The three statistical quantities~\cite{SCM1,SCM2,Simini,YY,Catt,Cir,Mat,GGG,SSS} characterizing the human
mobility dynamics are: ({\em i}) the total number $S(t)$
of distinct locations that an individual visited within time $t$,
({\em ii}) the probability $P(z)$ for an individual to visit the $z$th
new location, if he/she has already visited $z-1$ distinct locations,
and ({\em iii}) the fraction $P(k)$ of locations that have been
visited $k$ times. The quantity $P(z)$ can be used to infer whether
previously visited locations are more likely to be visited than newly
discovered locations, which we will show possesses a more complex form
than the well-known Zipf's law~\cite{Sim}. The quantity $P(k)$
is similar to the degree distribution in complex
networks~\cite{New,Al,AL}. The three quantities can
be used to validate our model through a detailed comparison
between theoretical prediction and numerical results.

\subsection{Analytical solutions.} We aim to obtain the analytic expectation values of the three
characterizing quantities. Since walkers are independent of each other,
it suffices to analyze a single walker.

(i) \emph{The number of distinct locations, $S(t)$.} $S(t)$ is defined as the total number of distinct positions visited by the person within time $t$. Inspired by the master equation method, we write down the probability of visiting a new position:
$$P_{\rm new}=\frac{M-S}{M+\lambda t}.\eqno{(1)}$$
By solving it, we have
$$S(t)=M-(M-1)\bigg (\frac{M+\lambda}{M+\lambda t}\bigg)^{\frac{1}{\lambda}}.\eqno{(2)}$$

(ii) \emph{The visit probability of positions discovered at different time, $P(z)$.} By using same method as above we have
$$P(z) =\frac{(M+\lambda t)(\lambda+1)}{(\lambda+N)\lambda t}\bigg(\frac{M-z}{M-1}\bigg)^{\lambda}.\eqno{(3)}$$

(iii) \emph{The visit probability of each position, $P(k)$.}
To calculate $P(k)$, we note that the total number of the visited locations
is $S$. Each location has its own ordinal $z$, which gives a relation between
$k$ and $z$. Suppose $k(z)$ is a monotonously decreasing function, we
can obtain its inverse $z(k)$, also a monotonously decreasing function.
The measure of $k=x$ is $|\Delta z|=|z'(k=x)\Delta k|$. We have

$$P(k=x)=\frac{|\Delta z|}{S}=\frac{|z^{'}(x)\Delta k|}{S}.\eqno{(4)}$$
As $k$ is an integer and $\Delta k=1$ in the system, we have
 $$P(k)=\frac{|z^{'}(k)|}{S(t)},\eqno{(5)}$$ where
 $$
\begin{cases}
|z^{'}(k)|=-z^{'}(k)=(M-1)\left[\frac{\lambda+M}{(M+\lambda t)(\lambda+1)}\right]^{\frac{1}{\lambda}}(k\lambda+1)^{\frac{1}{\lambda}-1},
\\
S(t)=M-(M-1)\left(\frac{M+\lambda}{M+\lambda t}\right)^{\frac{1}{\lambda}}.
\end{cases}\eqno{(6)}
\\
$$
More details can be seen in the appendix part.

\section{Results}

\subsection{Numerical validation on artificial systems.} We conduct systematic numerical simulations of our MPRW model to obtain
the scaling laws governing the three characterizing quantities, using
the concrete setting where 100 walkers are distributed in a space of
1000 locations and perform 1000 walks, i.e., $N=100$, $M=1000$ and $t=1000$.
As $S(t)$, $P(z)$, $P(k)$ are defined for each walker, it is necessary
to aggregate the results from all walkers to uncover the general features.
Our approach is the following. ({\em i}) For each user $i$, we obtain
the relation between $S^i(t)$ and $t$, where $S^i(t)$ is the total number
of previously visited distinct locations within time $t$. We have
$S(t)=(1/N) \sum_iS^i(t)$. ({\em ii}) To calculate $P(z)$, we let
$P(z_\alpha^i)$ be the probability of $i$'s visiting the location
$\alpha$. Say $i$ has visited $z-1$ distinct locations before walking
into $\alpha$. The quantity $P(z_\alpha^i)$ is then the
fraction of times that walker $i$ visited $\alpha$, and
we have $P(z_\alpha)=(1/N)\sum_iP(z_\alpha^i)$.
({\em iii}) For $P(k)$, we note that, each
location $\alpha$ can be visited by different times for each walker. Let
$k_\alpha^i$ be the number of times that walker $i$ visited $\alpha$.
The aggregated frequency of visit to $\alpha$ is $k_\alpha=\sum_ik_\alpha^i$,
and $P(k)$ can be obtained through the histogram of the sequence
$\{k_1, k_2, k_3, ..., k_M\}$.

\begin{figure}[h]
\centering
\includegraphics[width=\linewidth]{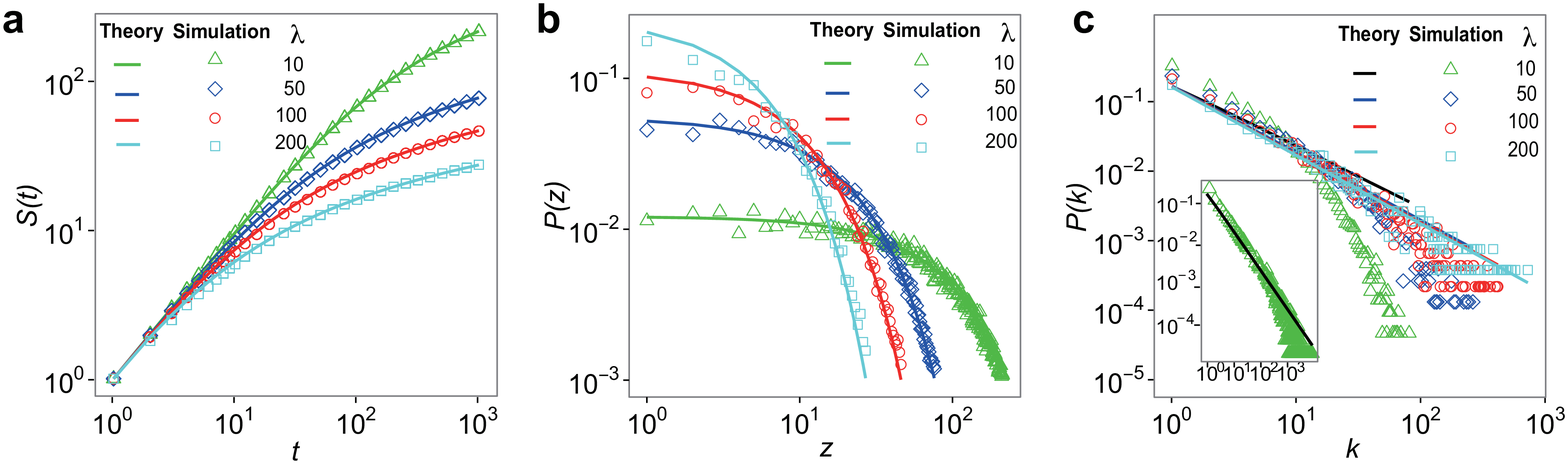}
\caption{\textbf{Comparison between simulation and analytical predictions
of MPRW model.} (a-c) For four values of the memory factor parameter
$\lambda$, the quantities $S(t)$, $P(z)$, and $P(k)$, respectively.
In (a) and (b), the agreement between simulation and theory is
remarkable, even for relatively short time duration $t = 1000$.
In (c), the theoretical prediction of $P(k)$ exhibits a power-law
scaling, but there are numerical deviations. The discrepancies can be
reduced by increasing the duration, as shown in the inset for $t = 10^5$.}
\label{fig:comparison}
\end{figure}

Fig.~\ref{fig:comparison}(a) shows the function $S(t)$ for different values
of $\lambda$, which is a sub-linear increasing function. We see
that a stronger memory effect corresponds to a smaller rate of increase,
which is natural due to individuals' resistance to explore new locations.
Fig.~\ref{fig:comparison}(b) shows the behavior $P(z)$, where we see
that the memory effect in general decreases the value $z$ at which $P(z)$
begins to decrease rapidly, meaning that nostalgic individuals tend to
discover few locations, a behavior that is consistent with that in
Fig.~\ref{fig:comparison}(a). For both Figs.~\ref{fig:comparison}(a) and
\ref{fig:comparison}(b), the simulation results agree well with the
analytical prediction. Fig.~\ref{fig:comparison}(c) shows the distribution
function $P(k)$, which exhibits a general power-law scaling behavior. For
large values of $\lambda$, the scaling exponent is about $-1$. However,
for small values of $\lambda$, $P(k)$ apparently deviates from the
theoretically predicted power-law form. The deviation is a result of
relatively short simulation duration. When we increase the duration to
$t=10^5$, the deviation diminishes, as shown in the inset of
Fig.~\ref{fig:comparison}(c).

\begin{figure}[h]
\centering
\includegraphics[width=\linewidth]{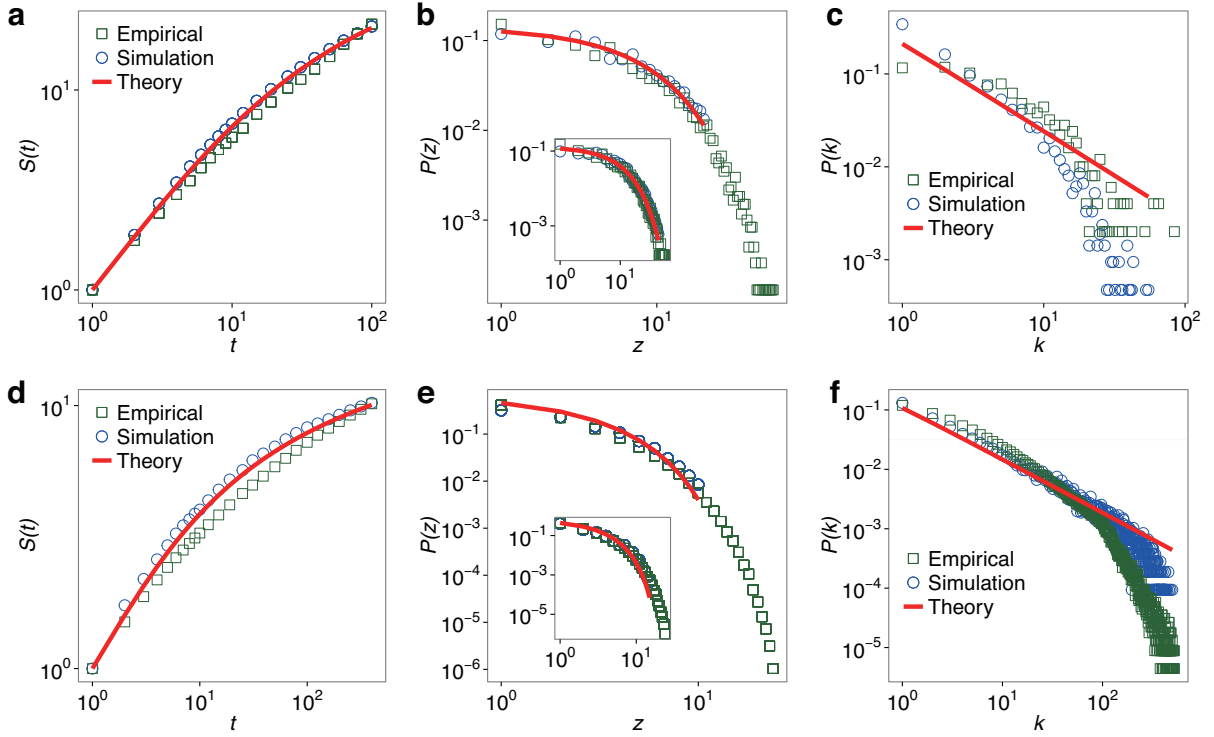}
\caption{\textbf{Empirical data analysis: check-ins and online-shopping
systems.} (a-c) From a mobile phone check-ins data set in New York city, the
quantities $S(t)$, $P(z)$, and $P(k)$, respectively, where the optimal memory
factor parameter is chosen to be $\lambda=23$. In (b), the $P(z)$ curve
from the simulated data has a shorter tail than that from the real data,
which can be corrected by extending the time duration from $t=100$ to
$t=10000$ in the MPRW model (inset). (d-f) The corresponding results
from a big online-shopping data set for $\lambda=10$.}
\label{fig:real_data}
\end{figure}

\subsection{Validation with the  location-based check-ins application.} We validate our model with  location-based check-ins data~\cite{LJJ}. Our
data set recorded, in New York city, the positions of 42035 individuals
as they use the check-in application, where the whole city is divided into 197 blocks. In order
to obtain long enough time series, we focus on the individuals who
have at least $100$ recorded locations and analyze their first $100$
records. There are 60 individuals whose recorded data fulfill this
requirement. The quantities $S(t)$, $P(z)$ and $P(k)$ are computed by
aggregating the data from different individuals. The parameters that can
be input to the MPRW model are thus $N=60$, $M=197$ and $t=100$. A key
to validating the model is the choice of some suitable value of the
memory factor, $\lambda$. The optimal value, denoted as $\lambda^*$,
can be estimated by comparing $S(t)$ from simulation and from real data.
Specifically, limiting the choices of $\lambda$ to integer values, we
can calculate a set of square distance values, $d(\lambda)$, between
the two $S(t)$ curves. The value of $\lambda^*$ is one that minimizes
$d(\lambda)$. For the mobile phone data, we have $\lambda^*=23$,
as shown in Fig.~\ref{fig:real_data}(a). We see that, for this
choice of $\lambda$, the model predicted function $S(t)$ agrees
well with that from the mobile phone check-ins data. With
$\lambda^*$ determined solely from $S(t)$, we also obtain a good
agreement between the model and empirical results for the
quantities $P(z)$ and $P(k)$, as shown in Figs.~\ref{fig:real_data}(b)
and \ref{fig:real_data}(c), respectively. which is remarkable. From
Fig.~\ref{fig:real_data}(b), we note that the $P(z)$ curve from the
model has a shorter tail than that from the real data. This is due to
the difference in the time scale between the simulation and real data,
e.g., $t=100$ in the real data may correspond to a much longer time
duration in the model. Extending the time duration to $t=10000$ in the
model gives a better agreement, as shown in the inset of
Fig.~\ref{fig:real_data}(b). We find that $S(t)$ and $P(k)$ are
immune to this effect, insofar as the time duration is not too small.

\subsection{Online commercial data.} The big data set is from \emph{Taobao.com}~\cite{ZDZ}. As a main
business branch of the Alibaba Group (a giant Chinese Internet company),
Taobao is regarded as China's equivalent of eBay. The data set consists
of the click records of Taobao users. When a user intends to make a purchase
on Taobao, he/she clicks a sequence of links to obtain the relevant
information (e.g., brand and price) of the product and then chooses
one to buy. This process can be regarded as users surfing online web
pages, i.e., movements in the cyberspace. Our data set consists of the
records of 34330 individuals. After initial filtering to remove the
individuals with abnormally long or very short click strings, we obtain a
slightly smaller data set with 33462 users, for which the total number
of web pages is $25$.

To cast the online-shopping process in the framework of MPRW, we regard
each web page as a location. To be consistent with the data, we set $N=100$,
$M=25$ and $t=500$. Using the same method as for the mobile phone data,
we determine the optimal value of the memory-factor parameter to be
$\lambda^*=10$. The results of $S(t)$, $P(z)$, and $P(k)$ are shown in
Figs.~\ref{fig:real_data}(d-f), respectively. Again, the results from
MPRW model agree well with those from the data [for $P(z)$ a good agreement
is achieved when an extended time duration, $t = 5000$, is used in the
model, as shown in the inset in (b)], suggesting the model's universal
applicability.

\section{Discussion}
To summarize, we develop a random walk model with a single
parameter to reproduce the statistical scaling behaviors of the three
quantities characterizing human mobility. The key element that makes
our model distinct from previous ones is a memory-preferential mechanism in limited space.
We demonstrate that, when this mechanism is incorporated into a standard
random walk process, the analytically predicted behaviors agree, at
a detailed and quantitative level, with those from two representative
real data sets, one for real world and another for cyberspace movements.
This is remarkable, considering that model is minimal with only a single
adjustable parameter, the memory-factor parameter. The main message is then
that, while various mechanisms can be considered for human mobility, such
as planted or social events~\cite{F-C} and gender
difference~\cite{G-S,K-L,J-E}, our findings
provide strong evidence that random walk with memory is the universal
underpinning of the human movement dynamics. While we assume in the
present work that the walkers are homogeneous, the analysis can be
extended to models incorporating memory heterogeneity.

Given a data set from any generic behavior of human movements, the
optimal memory-factor parameter for the MPRW model can be estimated
by comparing the behavior of an elementary statistical quantity from
data and model. This feature has the additional benefit of assessing
and quantifying the degree of intrinsic memory effect in the real
system, which has potential applications to problems of significant
current interest such as traffic optimization and online recommendation.

\section*{Acknowledgements} This work is supported by National Science Foundation of China (Grant No. 61573064, 61074116 and 11547188), the Youth Scholars Program of Beijing Normal University (grant No. 2014NT38), and the Fundamental Research Funds for the Central Universities Beijing Nova Programme, China. XYY acknowledges the support from the National Natural Science Foundation of China (Grant No. 61304177) and the Fundamental Research Funds of BJTU (Grant No. 2015RC042).

\addcontentsline{toc}{section}{Appendix. The derivation details of the analytical solutions}

\section*{Appendix. The derivation details of the analytical solutions}
\subsection*{The number of distinct locations, $S(t)$.}
We have the probability of visiting a new position Eq. (1). So we have the equation
$$\frac{dS}{dt}=\frac{M-S}{M+\lambda t}.\eqno{(7)}$$
By solving it, we have
$$S(t)=M-c\left(\frac{1}{M+\lambda t}\right)^{\frac{1}{_{\lambda}}},\eqno{(8)}$$
where $c$ is a constant. At the beginning of the random-walk, we have the initial condition: $t=1$ and $S(t)=1$. Accordingly, we can obtain the value of $c$ as
$$c=(M+\lambda)^{\frac{1}{\lambda}}(M-1).\eqno{(9)}$$
Thus we can have the solution of $S(t)$ as Eq. (2).

\subsection*{The visit probability of positions discovered at different time, $P(z)$.}
 Let $k_{z}$ denote the position $z$ has been visited $k$ times and $t_{z}$ denote the first time when position $z$ is visited.By using the same method, to $k_{z}$,we have
$$\frac{dk_{z}}{dt}=\frac{1+\lambda k_{z}}{M+\lambda t}. \eqno{(10)}$$
Solving it with the initial condition $t=t_{z}$ and $k_{z}=1$, we have
$$k_{z}=\frac{(\lambda+1)(\lambda t+M)}{\lambda (M+\lambda t_{z})}-\frac{1}{\lambda}.\eqno{(11)}$$
Actually, we have a hidden condition $S(t_{z})=z$. By using it, we can get
$$k_{z}=\frac{(M+\lambda t)(\lambda+1)}{(\lambda+M)\lambda}\bigg(\frac{M-z}{M-1}\bigg)^{\lambda}-\frac{1}{\lambda}.\eqno{(12)}$$
The visited frequency of each position $P(z)$ is proportional to $k_{z}$, so
 $$P(z) \propto k_{z}.\eqno{(13)}$$
As the sum of the $k$ equals to evolving time $t$, so we have
 $$P(z) =\frac{k_{z}}{t}\approx \frac{(M+\lambda t)(\lambda+1)}{(\lambda+N)\lambda t}\bigg(\frac{M-z}{M-1}\bigg)^{\lambda}.\eqno{(14)}$$

\subsection*{The visit probability of each position, $P(k)$.}We already have $S(t)$ as Eq.(2). And we can solve $k(z)$ inversely to obtain
$$z(k)=M-(M-1)\bigg[\frac{(\lambda k+1)(\lambda+M)}{(\lambda+1)(M+\lambda t)}\bigg]^{\frac{1}{\lambda}}.\eqno{(15)}$$
So we have Eq. (6) and $P(k)$ as Eq. (5).

\end{document}